\documentclass[showpacs,twocolumn,superscriptaddress]{revtex4}

\usepackage{doi}%<----------
\usepackage{hyperref}
\hypersetup{
	colorlinks=true,        % false: boxed links; true: colored links
	linkcolor=blue,         % color of internal links
	citecolor=cyan,         % color of links to bibliography
}

\usepackage{graphicx}% Include figure files
\usepackage{dcolumn}% Align table columns on decimal point
\usepackage{bm}% bold math
\usepackage{color}

\usepackage{amsmath}
\usepackage{amssymb}

\def\a{\alpha}
\def\r{\rho}
\def\s{\sigma}
\def\t{\tau}
\def\m{\mu}
\def\n{\nu}
\def\k{\kappa}
\def\th{\theta}
\def\g{\gamma}\def\G{\Gamma}
\def\L{\Lambda}\def\l{\lambda}
\def\D{\Delta}
\def\la{\langle}
\def\ra{\rangle}
\def\o{\omega}\def\O{\Omega}
\def\d{\delta}
\def\p{\partial}

\def\half{\textstyle{\frac{1}{2}}}

\def\bdoc{\begin{document}}
	\def\edoc{\end{document}}
\def\beq{\begin{equation}}
\def\eeq{\end{equation}}
\def\bea{\begin{eqnarray}}
\def\eea{\end{eqnarray}}
\def\ben{\begin{enumerate}}
	\def\een{\end{enumerate}}
\def\la{\langle}
\def\ra{\rangle}
\def\a{\alpha}
\def\b{\beta}
\def\g{\gamma}
\def\G{\Gamma}
\def\d{\delta}
\def\D{\Delta}
\def\e{\epsilon}

\def\th{\theta}
\def\k{\kappa}
\def\l{\lambda}
\def\m{\mu}
\def\n{\nu}
\def\o{\omega}
\def\p{\pi}
\def\r{\rho}
\def\s{\sigma}
\def\t{\tau}
\def\L{{\cal L}}
\def\S{\Sigma }
\def\gsim{\; \raisebox{-.8ex}{$\stackrel{\textstyle >}{\sim}$}\;}
\def\lsim{\; \raisebox{-.8ex}{$\stackrel{\textstyle <}{\sim}$}\;}
\def\gtrsim{\gsim}
\def\lessim{\lsim}
\def\loc{{\rm local}}
\def\vm{v_{\rm max}}
\def\bh{\bar{h}}
\def\del{\partial}
\def\nab{\nabla}
\def\half{{\textstyle{\frac{1}{2}}}}
\def\fourth{{\textstyle{\frac{1}{4}}}}

\def\bD{{\bf D}}
\def\bE{{\bf E}}
\def\bF{{\bf F}}
\def\bB{{\bf B}}
\def\bP{{\bf P}}
\def\bV{{\bf v}}
\def\bv{{\bf v}}
\def\bx{{\bf x}}
\def\by{{\bf y}}
\def\bz{{\bf z}}
\def\ba{{\bf a}}
\def\bd{{\bf d}}
\def\bs{{\bf s}}
\def\bn{{\bf n}}
\def\bp{{\bf p}}

\def\O{\Omega}

\def\br{{\bf r}}
\def\bnab{{\bf \nab}}

\def\tE{\tilde{E}}
\def\tL{\tilde{L}}

\begin{document}
	
	\title{%Motion of Test Particles and their 
	Dynamics and Fundamental Frequencies of Test Particles Orbiting Kerr-Newman-NUT-Kiselev Blacks Hole in Rastall Gravity}
	
	\author{Bakhtiyor Narzilloev}
	\email[]{nbakhtiyor18@fudan.edu.cn}
	\affiliation{Center for Field Theory and Particle Physics and Department of Physics, Fudan University, 200438 Shanghai, China }
	\affiliation{Akfa University, Kichik Halqa Yuli Street 17, Tashkent 100095, Uzbekistan}
	\affiliation{Ulugh Beg Astronomical Institute, Astronomy St.  33, Tashkent 100052, Uzbekistan}

	\author{Ibrar Hussain}
	\email[]{ibrar.hussain@seecs.nust.edu.pk}
	
	\affiliation{School of Electrical Engineering and Computer Science, National University of Sciences and Technology, H-12, Islamabad, Pakistan}
	
	\author{Ahmadjon~Abdujabbarov}
	\email[]{ahmadjon@astrin.uz}
	\affiliation{Shanghai Astronomical Observatory, 80 Nandan Road, Shanghai 200030, P. R. China}
	\affiliation{Ulugh Beg Astronomical Institute, Astronomy St.  33, Tashkent 100052, Uzbekistan}
	\affiliation{Institute of Nuclear Physics, Ulugbek 1, Tashkent 100214, Uzbekistan}
	\affiliation{National University of Uzbekistan, Tashkent 100174, Uzbekistan}
	\affiliation{Tashkent Institute of Irrigation and Agricultural Mechanization Engineers, Kori Niyoziy 39, Tashkent 100000, Uzbekistan}

	\author{Bobomurat Ahmedov}
	\email[]{ahmedov@astrin.uz}
	
	\affiliation{Ulugh Beg Astronomical Institute, Astronomy St.  33, Tashkent 100052, Uzbekistan}
	\affiliation{National University of Uzbekistan, Tashkent 100174, Uzbekistan}
	\affiliation{Tashkent Institute of Irrigation and Agricultural Mechanization Engineers, Kori Niyoziy 39, Tashkent 100000, Uzbekistan}
	
	\author{Cosimo Bambi}
	\email[]{bambi@fudan.edu.cn}
	\affiliation{Center for Field Theory and Particle Physics and Department of Physics, Fudan University, 200438 Shanghai, China }

	\date{\today}
	\begin{abstract}
		The spacetime properties in the exterior of the Kerr-Newman-NUT-Kiselev black hole in the Rastall theory of gravity, through particle dynamics are investigated with the aim to find possible degeneracy of the different black hole parameters. We show that the effective potential, the energy, and the angular momentum of a test particle moving in the spacetime of such a black hole strongly depend on the central black hole parameters. We also evaluate the innermost stable circular orbit radii of test particles and show how the spacetime parameters can act on them. Further, we show the results for the fundamental frequencies of test particles moving at small distances from the circular orbits in the equatorial plane. We demonstrate that change in the Rastall parameter $\kappa \lambda$, can make the radial epicyclic frequency to become zero at larger distances from the central source. We also notice that for the vertical epicyclic frequencies, in the case of the Kerr-Newman-NUT-Kiselev black hole in the Rastall theory of gravity lower frequencies are observed as compared to the frequencies observed in the case of the Kerr black hole. Finally, we show that for the Kerr-Newman-NUT-Kiselev black hole in the Rastall gravity, the Keplerian frequencies are almost identical with the frequencies noticed in the case of the Kerr black hole and the difference between the two can only be observed in the regions in a very close vicinity to the central black hole, studied in the present work.
	\end{abstract}
	\maketitle
	
	%\section{Introduction}
	
	\section{Introduction}
	\label{sec1}
The study of motion of test particles in black hole (BH) close environment has its own importance as it may be helpful in understanding the different properties of BHs in the strong gravitational field regime. The resent observation of gravitational waves produced by a binary system of two BHs, detected by the LIGO, \cite{1}, the first ever image of the astrophysical BH $M87$ revealed by the Event Horizon Telescope (EHT) collaboration \cite{2} and the very recent observation of the X-rays released by a supermassive BH at the center of a galaxy \cite{2a}, have further increased the interest of researchers in probing BH Physics in the strong gravitational field regime. Theoretically BHs are exact and singular solutions of the Einstein Field Equations (EFEs) of the General Theory of Relativity (GTR) \cite{3}. These BHs are completely characterised by only three parameters, namely, mass, rotation or spin and charge.\\      

 The presence of plasma or dust particles in the vicinity of BHs accreting on them is an appealing candidate to investigate circular time like geodesics of both neutral and charged particles in such spacetimes. Charged and neutral particle dynamics in charged/uncharged BH spacetimes with and without rotation in the GTR have extensively been studied by different authors \cite{4,5,6,7,9,10}. In the literature the effects of different fields like magnetic field and quintessence field, a potential candidate for the dark energy which is assumed to be responsible for the current accelerated expansion of our Universe, have been analysed on the motion of particles in the spacetimes of different BHs \cite{11,12,13,Narzilloev20c,15,16,17}. Circular geodesics in spacetimes of BHs in theories of gravity, other than GTR, have also been investigated in the literature, see e.g, \cite{Narzilloev:2020b, Narzilloev19, Narzilloev20a, Narzilloev20b, Narzilloev20c, Hakimov17, Narzilloev21, Narzilloev21a, Narzilloev21b}.\\
 
 The Kerr solution of the EFEs, represents the gravitational field of a rotating BH. It is thought that astrophysical black holes have no charge and hence are electrically neutral, therefore, the Kerr solution can be considered as the most suitable candidate to describe an astrophysical BH. However, there is a charged generalization of the Kerr solution in the literature, known as the Kerr-Newman solution of the Einstein-Maxwell field equations.\ \cite{18}. Charged particle motion in the spacetime of a charged and rotating black hole has been considered by different authors (see e.g. \cite{19,20,21,22,23,24}). The geodesic motion in the vicinity of Kerr-Newman BH has also been studied in the presence of magnetic field \cite{25}. A further generalization of the Kerr-Newman solution has been obtained with the NUT parameter \cite{26}, and the study of the dynamics of test particles in this spacetime has been carried out in the literature to investigate the effects of the NUT parameter on the particle motion \cite{27,22,29}. Kiselev has presented a BH solution in the presence of dark energy for a point gravitating source \cite{30}. A charged \cite{29}, and then a rotating version of the Kiselev BH has been obtained in the literature \cite{32,33}, and studies of the geodesic motion have been done in these spacetimes \cite{34,35,36,17}. In a recent work the Kerr-Newman BH with the NUT parameter in the presence of the dark energy has been obtained in the Rastall theory of gravity (RG), known as the Kerr-Newman-NUT-Kiselev (KNNK) BH \cite{38}. The RG is a modification of the GTR, where a non-minimal coupling between geometry and matter fields is taken into account. The usual conservation law of the energy–momentum tensor $T^{\mu \nu}$, is not respected in the theory of RG. For further details on the RG one may see (e.g. \cite{39,40}). Here we are interested to investigate test particle motion in the spacetime of charged and rotating black hole, with the NUT parameter in the presence of the quintessence field in the theory of RG, to examine that how the circular orbits of a test particle can be influenced in such a spacetime by the BH parameters and as well as by the NUT and the Rastall parameters in the presence of the quintessence field.\\   
 
 Another interesting astrophysical phenomenon to be studied for test particles, close to the stable circular orbits is the fundamental frequencies of the quasi periodic oscillations (QPOs) \cite{41}. These QPOs are actually detected in the X-ray radiation of a binary systems of BHs surrounded by some accretion disc of some matter that flowing from companion stars. It is assumed that in the circular discs close to the innermost stable circular orbits (ISCOs), friction is so strong that the particles in such circular discs starts to emit X-rays \cite{42}. The QPOs in the X-ray radiation are of interest to astrophysicists as they may be helpful in the accurate measurement of the mass, charge, and spin of BHs. The spectroscopy techniques (the frequency distribution of photons), and timing (photon number time dependence), can be used to extract useful information from a particular source \cite{43}. To examine and get insight in the strong fields due to gravity, the fundamental frequencies of the QPOs in the X-rays from the accretion disc of matter around BHs, have been studied in the recent literature \cite{44,45,46,47}. Some different models including the disc-seismic model, the hot-spot model, the resonance model and the warped disc model have been proposed for the comprehension of QPOs, in the literature \cite{41}. So far no exact mechanism is known for the production of the QPOs detected in the X-rays, thus none of the above referred models can be fitted to the observational data from different astrophysical sources \cite{49}. 

In the present work we are keen to investigate the fundamental frequencies of test particles in the vicinity of circular orbits around the KNNK BH in the theory of RG, to see the effects of the involved spacetime parameters on them. We structured our paper as follows: In Sec.~\ref{sec2} we briefly introduce the KNNK spacetime in the RG. In Sec.~\ref{sec3} we study particle motion around the KNNK BH. In Sec.~\ref{sec4} we focus on the properties of fundamental frequencies of test particles moving around the KNNK BH in the RG. 
It is worth to note that there have been proposed several different models to explain the nature of QPO objects. Here we are concentrated on those models where QPOs are explained using fundamental frequencies of test particles orbiting around compact objects. 
In Sec.~\ref{sec5} we make a conclusion of the results obtained in the work. We use natural system of units where $G=c=1$.
 
	\section{Kerr-Newman-NUT-Kiselev spacetime in the Rastall gravity}\label{sec2}
	In this section we briefly introduce the spacetime around a BH described by the KNNK metric in the theory of the RG, that has the following form (see \cite{Sakti2019krw})
	
	\begin{eqnarray}
	ds^2&=&-\frac{\Delta}{\rho^2} [dt-\{a \sin^2\theta+2 l (1-\cos\theta)\} d\phi]^2\\\nonumber
	&+&\frac{\rho^2}{\Delta} dr^2+\rho^2 d\theta^2+\frac{\sin^2\theta}{\rho^2} [a dt-\{r^2+(a+l)^2\} d\phi]^2\, ,
	\end{eqnarray}
	where 
	\begin{eqnarray}
	\Delta&=&r^2-2 M r +a^2+e^2+g^2-l^2-\alpha r^v, \\ v
	&=&\frac{1-3 \omega}{1-3 \kappa \lambda (1+\omega)}, \\ \rho^2&=&r^2+(l+a\cos\theta)^2.
	\end{eqnarray}
	Here, $M$ and $a$ are the gravitational mass and spin of the BH respectively. The other parameters appearing in the metric of the KNNK BH in the RG are as follows: $l$ is the NUT parameter, $\alpha$ is the quintessential intensity, $\kappa\lambda$ is Rastall parameter, $\omega$ is the parameter of equation of state of the quintessence, $e$ and $g$ correspond to the electric and magnetic charge of BH, respectively \cite{Sakti2019krw}. One can introduce new parameter $q^2=e^2+g^2$ that involves the contribution of the electric and magnetic charges of a black hole to the given spacetime.
	
Depending on the value of the parameters $\omega$ and $\kappa\lambda$ the KNNK BH solution derived in RG has several horizons, which is determined as the coordinate singularity of the spacetime and is a null hypersurface of constant $r$. Accordingly, the horizons are determined as the roots of the following equation
\begin{eqnarray}
r^2-2Mr+a^2+q^2-l^2-\alpha r^v=0\ .
\end{eqnarray}
For selected values of $\omega$, $\kappa\lambda$, the horizon coincides to have only inner and outer parts or there is no cosmological horizon. In this specific case, the thermodynamic properties of BHs are easier to be explored. It is the reason why in \cite{Xu2019}, the authors consider only two roots for the horizon. When there are more than two roots for the horizon, it may lead to more complicated computation, for example, to study the entropy product. In \cite{Sakti2019krw}, it has been provided that under what circumstances there will be two analytic roots which are presented here in the Table \ref{tab1}. 
\begin{table}[h]\label{tab1}
\caption{\label{tab1}{Exact analytical expressions for inner and outer horizons  for selected values of parameters $\omega$ and $\kappa\lambda$.}}
\begin{tabular}{lll}
\hline\hline
$\omega$, $\kappa\lambda$ &  & Horizon ($r_\pm$) \\\hline\\
0,0 & & $\left(M+\frac{\alpha}{2}\right) \pm\sqrt{\left(M+\frac{\alpha}{2}\right)^2+l^2-a^2-q^2}$\\\\
-1/3,0 & & $\frac{M}{1-\alpha} \pm \frac{\sqrt{M^2-(a^2+q^2-l^2) (1-\alpha)}}{1-\alpha}$\\\\
0,1/6 & & $\frac{M}{1-\alpha} \pm \frac{\sqrt{M^2-(a^2+q^2-l^2) (1-\alpha)}}{1-\alpha}$\\\\
-1/3,-1/2 & & $\left(M+\frac{\alpha}{2}\right) \pm\sqrt{\left(M+\frac{\alpha}{2}\right)^2+l^2-a^2-q^2}$\\\\
1/3,0 & & $M \pm\sqrt{M^2+l^2+\alpha-a^2-q^2}$\\\hline\hline
\end{tabular}
\end{table}

%	\textcolor{blue}{We need to fulfil this part}

	\section{Test particle motion in the vicinity of the KNNK BH in the RG}\label{sec3}
\begin{figure*}[t!]
		\begin{center}
			\includegraphics[width=0.99\linewidth]{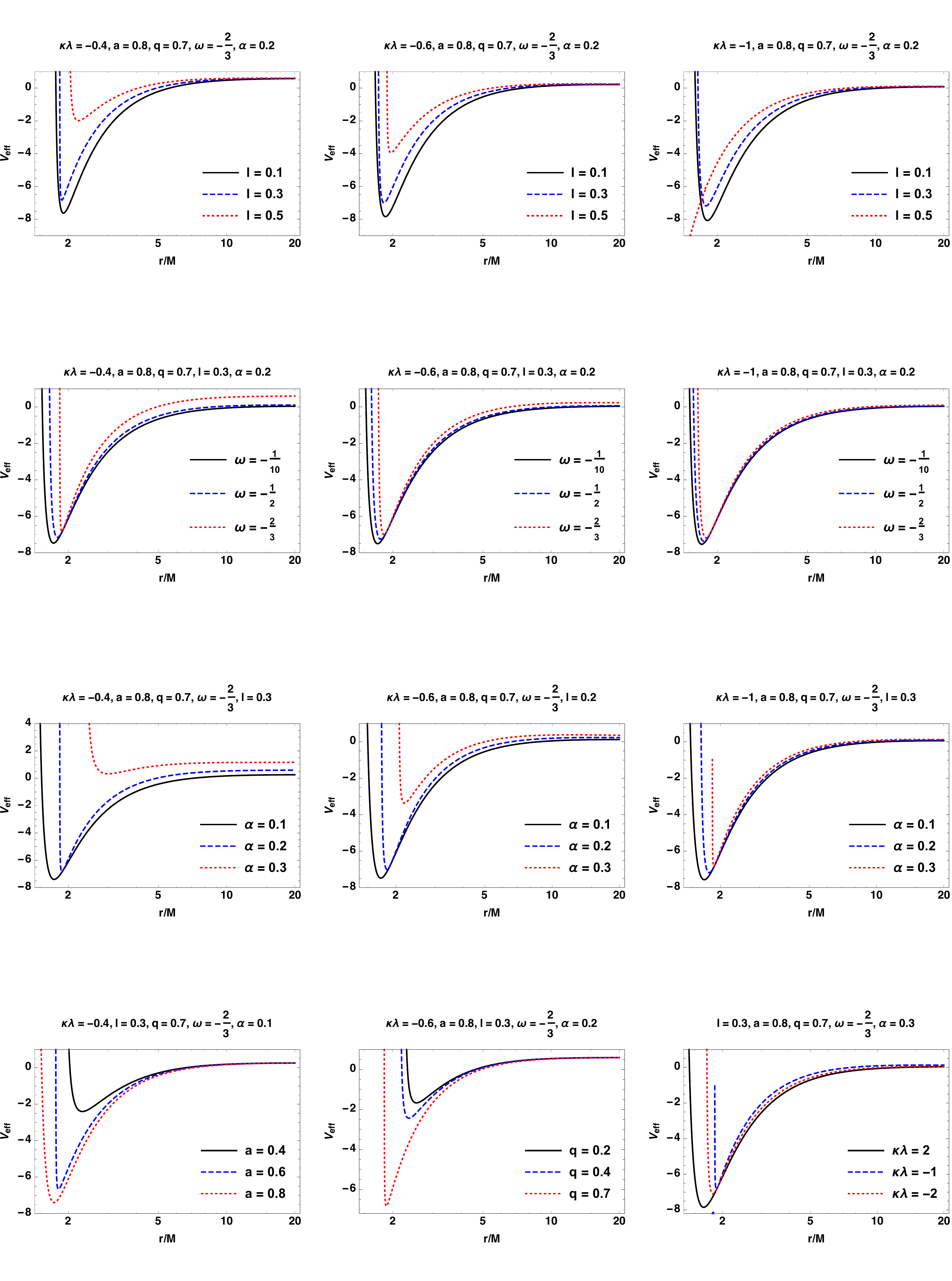}
		\end{center}
		\caption{The radial dependence of the effective potential of a test particle moving in the spacetime of the KNNK BH in RG for the fixed $\mathcal{E}=1$ and $\mathcal{L}=6$ and various values of the spacetime parameters. \label{Veff}}
	\end{figure*}

	One of the well known tool to test the spacetime around compact massive objects is the investigation of the dynamics of test particles moving around the central object. One can write the equation of motion of the test neutral particle in the following way
	\begin{eqnarray}\label{HJ}
	g^{\alpha\beta} \frac{\partial S}{\partial x^\alpha} \frac{\partial S}{\partial x^\beta}=-m^2\ ,
	\end{eqnarray}
	that corresponds to the Hamilton-Jacobi equation. One can solve this equation to get the following expression for the effective potential of the particle moving in the equatorial plane ($\theta=\pi/2$)
	\begin{eqnarray}
	V_{eff}(r)=\frac{\mathcal{E}^2 g_{\phi \phi}+2 \mathcal{E} \mathcal{L} g_{t \phi}+\mathcal{L}^2 g_{tt}}{g_{t\phi}^2-g_{tt}g_{\phi \phi}}-1\, ,
	\end{eqnarray}
	with the specific energy $\mathcal{E}=E/m$ and the specific angular momentum $\mathcal{L}=L/m$ of the test particle. %The explicit form of the effective potential is shown in Eq.\eqref{a1} in Appendix~\ref{app}. 
	For the chosen values of the parameters of the given spacetime one can plot the radial dependence of the effective potential as shown in Fig.~\ref{Veff}. It is worth noting that the given spacetime has many parameters and it is better to set up some strict combinations of these parameters to make the task easier. Throughout the paper we use the following combinations of the parameters $a=0.4, 0.5, 0.6, 0.8$; $q=0.2, 0.4, 0.7$; $\omega=-\frac{1}{10}, -\frac{1}{2}, -\frac{2}{3}$; $\kappa\lambda=-0.4, -0.6, -1, -2, 2$; $\alpha=0.1, 0.2, 0.3$ and $l=0.1, 0.2, 0.3, 0.5$. In the top panel of the Fig.~\ref{Veff} it has been shown how the effective potential changes with the change of the parameters $l$ and $\lambda$. One can see from the top panels of the same Fig. that increase of the NUT parameter increases the effective potential as well when other parameters are taken to be constant. It is also seen that increase of the parameter $\lambda$ slightly shifts the lines towards the smaller radii of the circular orbits. In the second row of the Fig.~\ref{Veff} similar scenario has been shown, where the lines for the different values of the parameter $\omega$ have taken place. It has been presented that decrease of the parameter $\omega$ makes the effective potential to take bigger values while the change of the parameter $\lambda$ shifts the lines towards each other. In the last plot of the second row it has been shown that for small values of the parameter $\lambda$, lines become very close to each other. In the third row of the Fig.~\ref{Veff} the behaviour of the effective potential for various values of the parameter $\alpha$ is illustrated. In the Fig.~\ref{Veff} it is clearly seen that with the increase of the parameter $\alpha$ the effective potential becomes bigger and the shape of the lines changes considerably with the decrease of the parameter $\lambda$. In the last row of the same Fig. several combinations of the spin $a$, charge $q$, and the parameter $\lambda$ has been presented. One can clearly see that increase of these three parameters reduces the effective potential. It is also seen from the last panel that in the large distances the effect of these parameters on the effective potential of the test particle becomes very weak.
%	\textcolor{blue}{Explanation of the figures needed to be added here}

\begin{figure*}[t!]
		\begin{center}
			\includegraphics[width=0.99\linewidth]{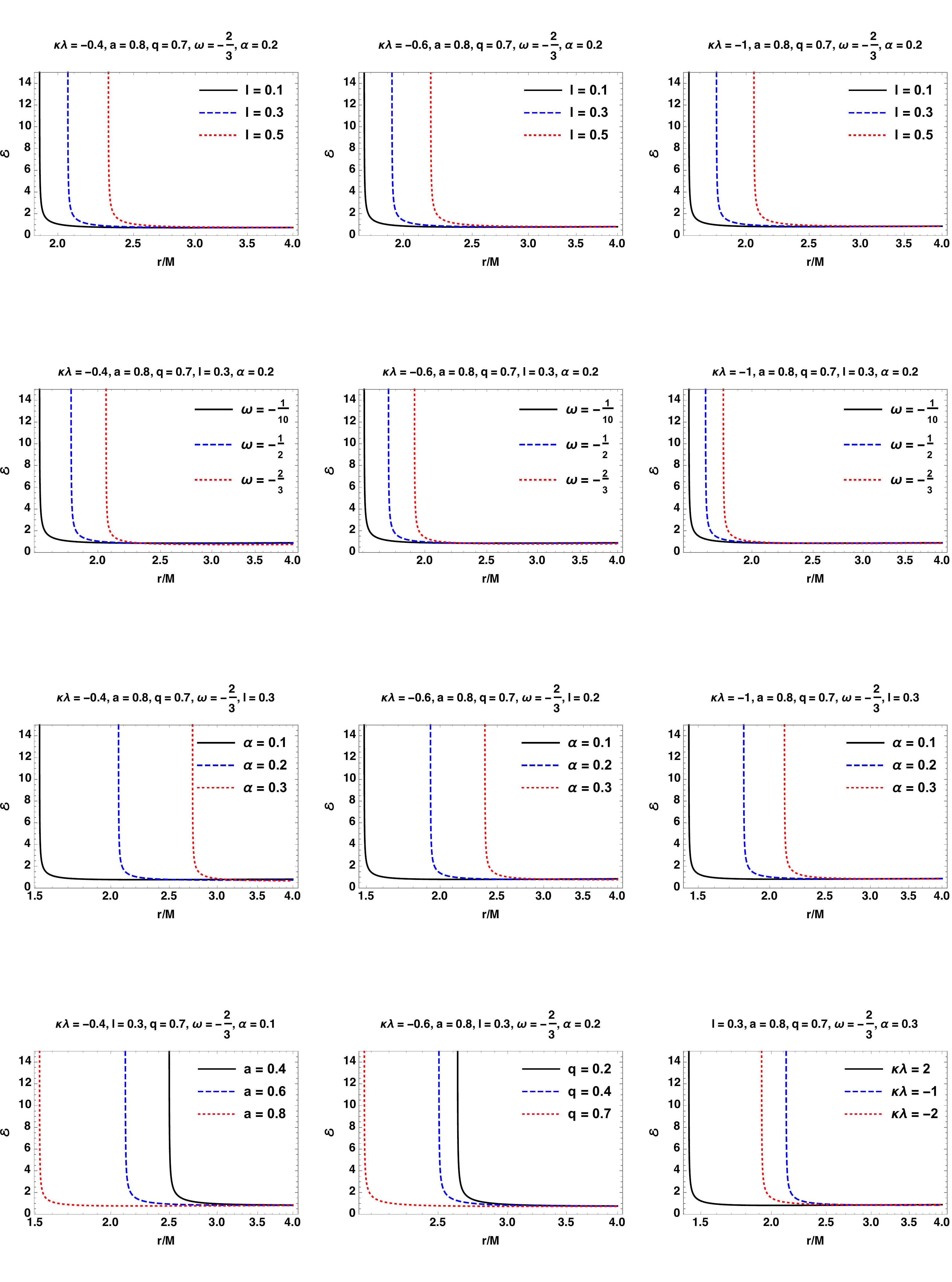}
		\end{center}
		\caption{The dependence of the energy of a test particle, moving in circular orbits around the KNNK BH in the RG, on the values of spacetime parameters. \label{E}}
	\end{figure*}

	For the particle to move in circular orbits one can set the following conditions on the effective potential
	\begin{eqnarray}
	V_{eff}(r)=0,\label{c1}\\
	V_{eff}'(r)=0,\label{c2}
	\end{eqnarray}
	that makes the energy and the angular momentum of the test particle to have the following form
	\begin{eqnarray}
	\mathcal{E}=\frac{-g_{tt}-\Omega g_{t\phi}}{\sqrt{-g_{tt}-2\Omega g_{t\phi}-\Omega^2 g_{\phi \phi}}}\ ,
	\end{eqnarray}
	and
	\begin{eqnarray}
	\mathcal{L}=\frac{\Omega g_{\phi \phi}+g_{t \phi}}{\sqrt{-g_{tt}-2\Omega g_{t\phi}-\Omega^2 g_{\phi \phi}}}\, ,
	\end{eqnarray}
	where
	\begin{eqnarray}
	\Omega=\frac{d\phi}{dt} = \frac{-g_{t\phi,r}\pm\sqrt{\{-g_{t\phi,r}\}^2-\{g_{\phi \phi, r}\} \{g_{tt,r}\}}}{g_{\phi \phi, r}}.
	\end{eqnarray}
%	gives the Keplerian frequency with the explicit form presented in Eq.\eqref{a2} in Appendix~\ref{app}.
	The dependence of the energy of the test particle on the radius of the circular orbit is plotted in Fig.~\ref{E}. Combinations of the spacetime parameters are taken to be the same as in the case of the plots for the effective potential discussed above. From the top panels of the Fig.~\ref{E} it can be seen that increase of the NUT parameter shifts the energy of the test particle moving on circular orbit towards the larger distances while the change of the parameter $\lambda$ affects negligibly small for the given combinations of the other parameters. From the second row of the Fig.~\ref{E} one can see that when one increases the value of the parameter $\omega$ the energy of the test particle shifts towards the central gravitating object described by KNNK spacetime in the RG. Similar to the effective potential one can see here that decrease of the parameter $\lambda$ makes the lines close to each other. In the third row of the Fig.~\ref{E} it is shown that increase of the parameter $\alpha$ shifts the lines to the right and the decrease of the parameter $\lambda$ reduces this shift. In the last row of the Fig.~\ref{E}  the change of the energy of the test particle with the change of the parameters $a$, $q$, and $\lambda$ is shown. From the same Fig. one can see the similar behaviour of the lines with the increase of these parameters shifting the lines to the left.

	% \textcolor{blue}{Explanation of the figures needed to be added here}
	
\begin{figure*}[t!]
		\begin{center}
			\includegraphics[width=0.99\linewidth]{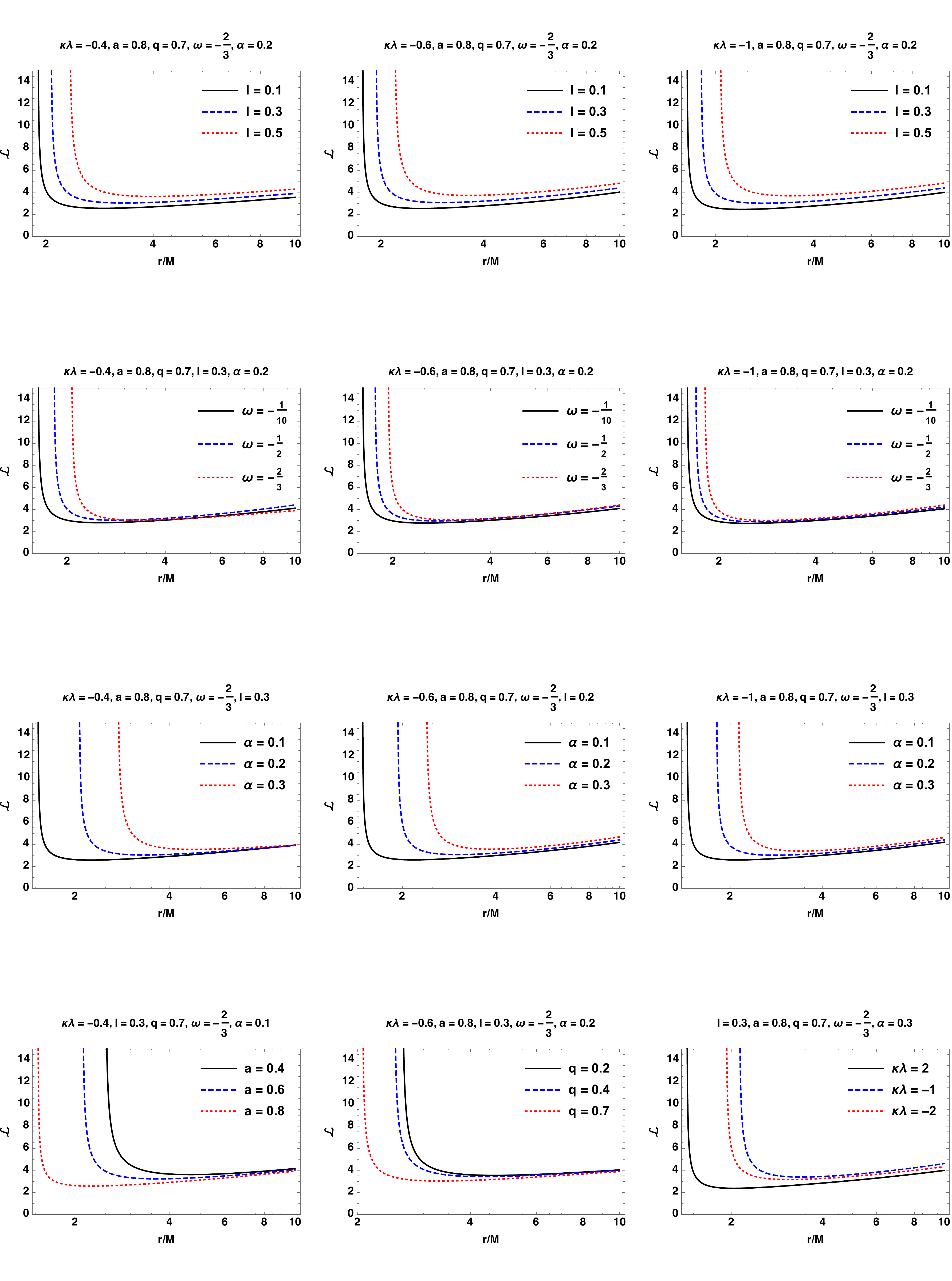}
		\end{center}
		\caption{The dependence of the angular momentum of a test particle making circular revolutions around the KNNK BH in the RG on the different values of the spacetime parameters. \label{L}}
	\end{figure*}

	One can also show the dependence of the angular momentum of the test particle from the radius of the circular orbits radius as shown in Fig.~\ref{L}. From the top panels it can be seen that increase of the NUT parameter makes the angular momentum of the test particle moving on circular orbits bigger while the decrease of the parameter $\lambda$ has negligible effect on it for the chosen combination of the other parameters. It has been shown in the second row of Fig.~\ref{L} that for the bigger values of the parameter $\omega$ the values of the angular momentum become smaller and the change of the parameter $\lambda$ slightly changes the shape of the lines. One can see from the third row of the Fig.~\ref{L} that the angular momentum becomes bigger for the bigger values of the parameter $\alpha$. In the last panel of the same Fig. one can see that increase of the parameters $a$, $q$, and $\lambda$ decrease the angular momentum of the test particle moving on the circular orbits around the KNNK BH in the RG.
	
	% \textcolor{blue}{Explanation of the figures needed to be added here}
	
\begin{figure*}[t!]
		\begin{center}
			\includegraphics[width=0.46\linewidth]{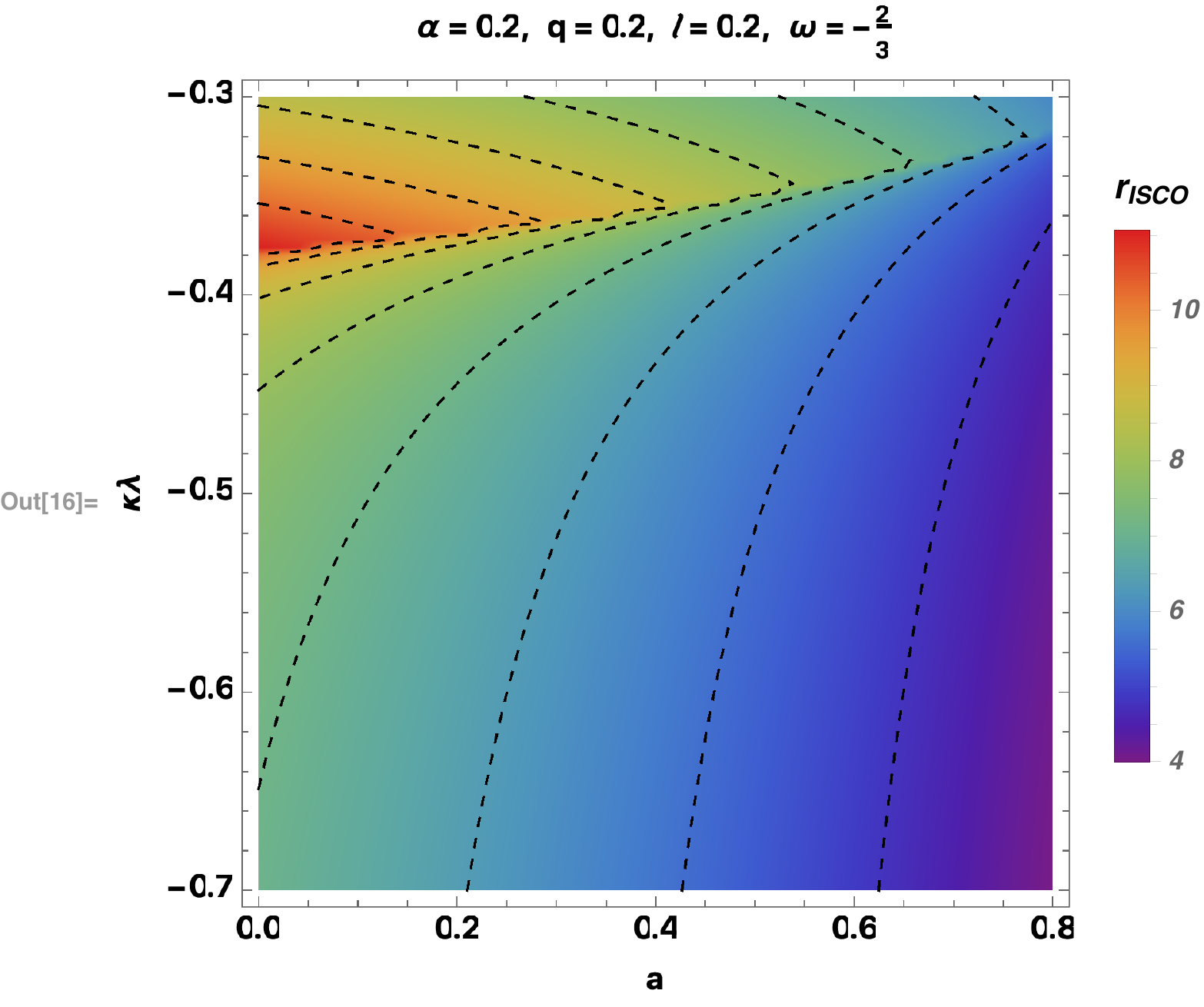}
			\includegraphics[width=0.46\linewidth]{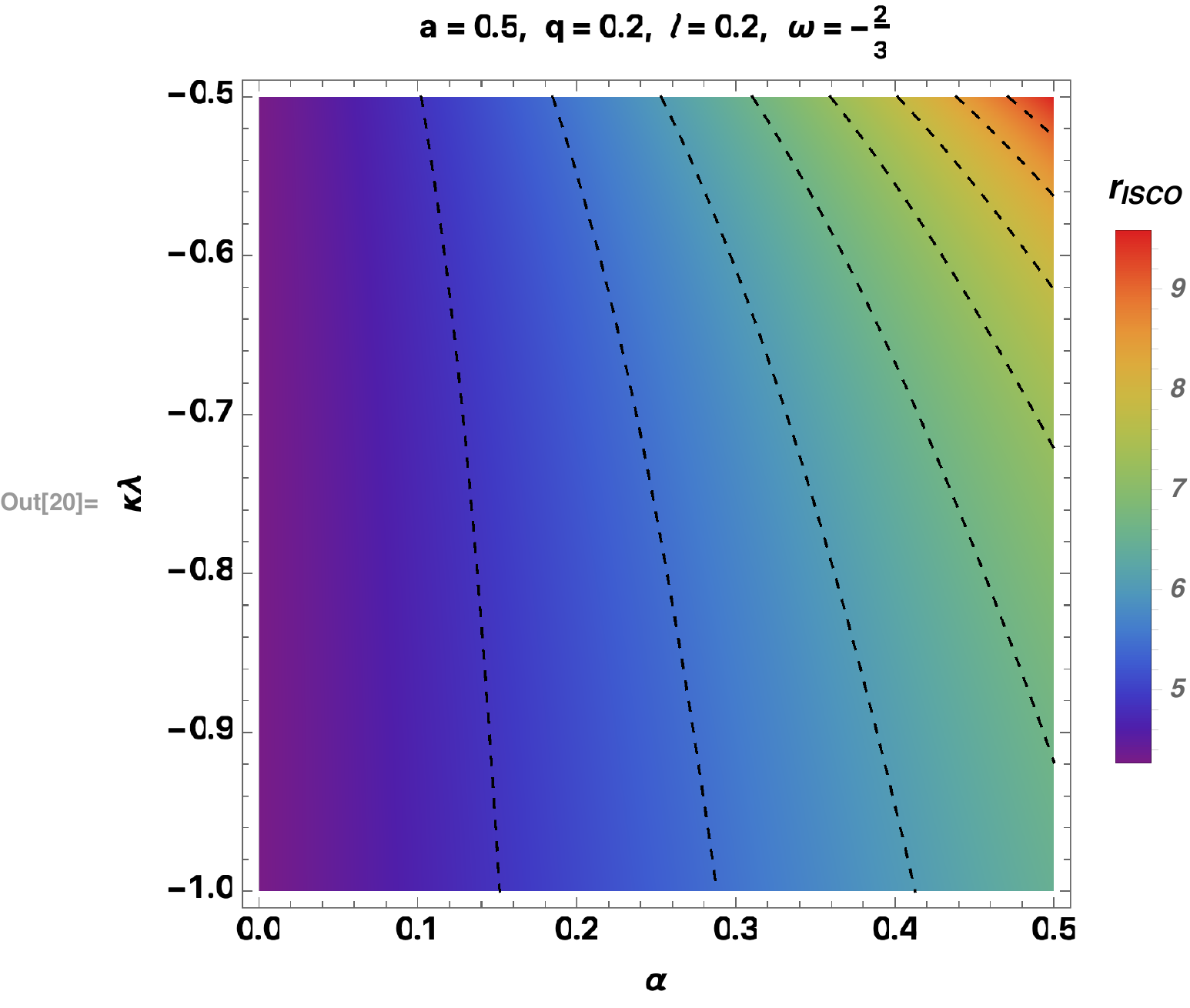}
			\includegraphics[width=0.45\linewidth]{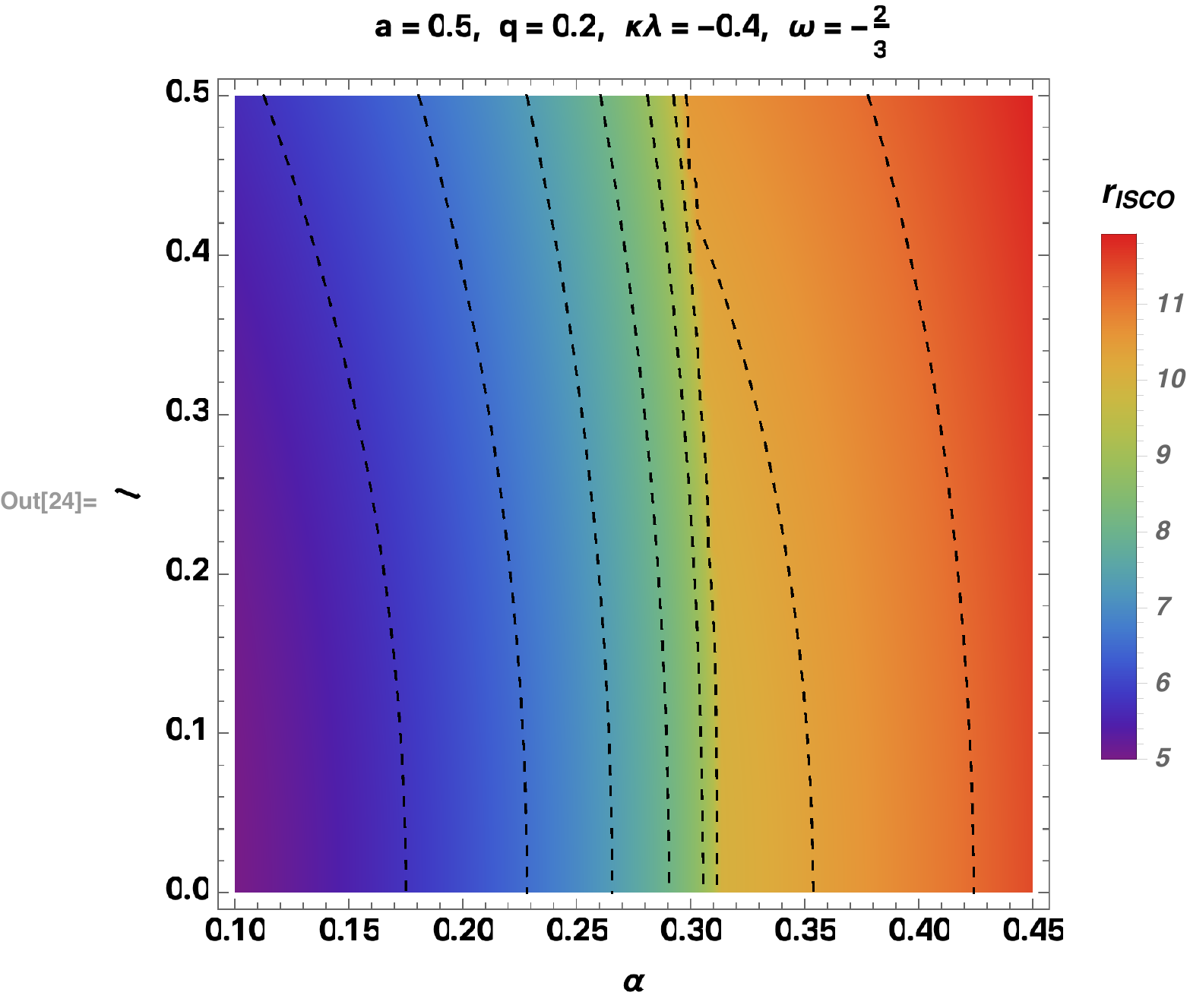}
			\includegraphics[width=0.45\linewidth]{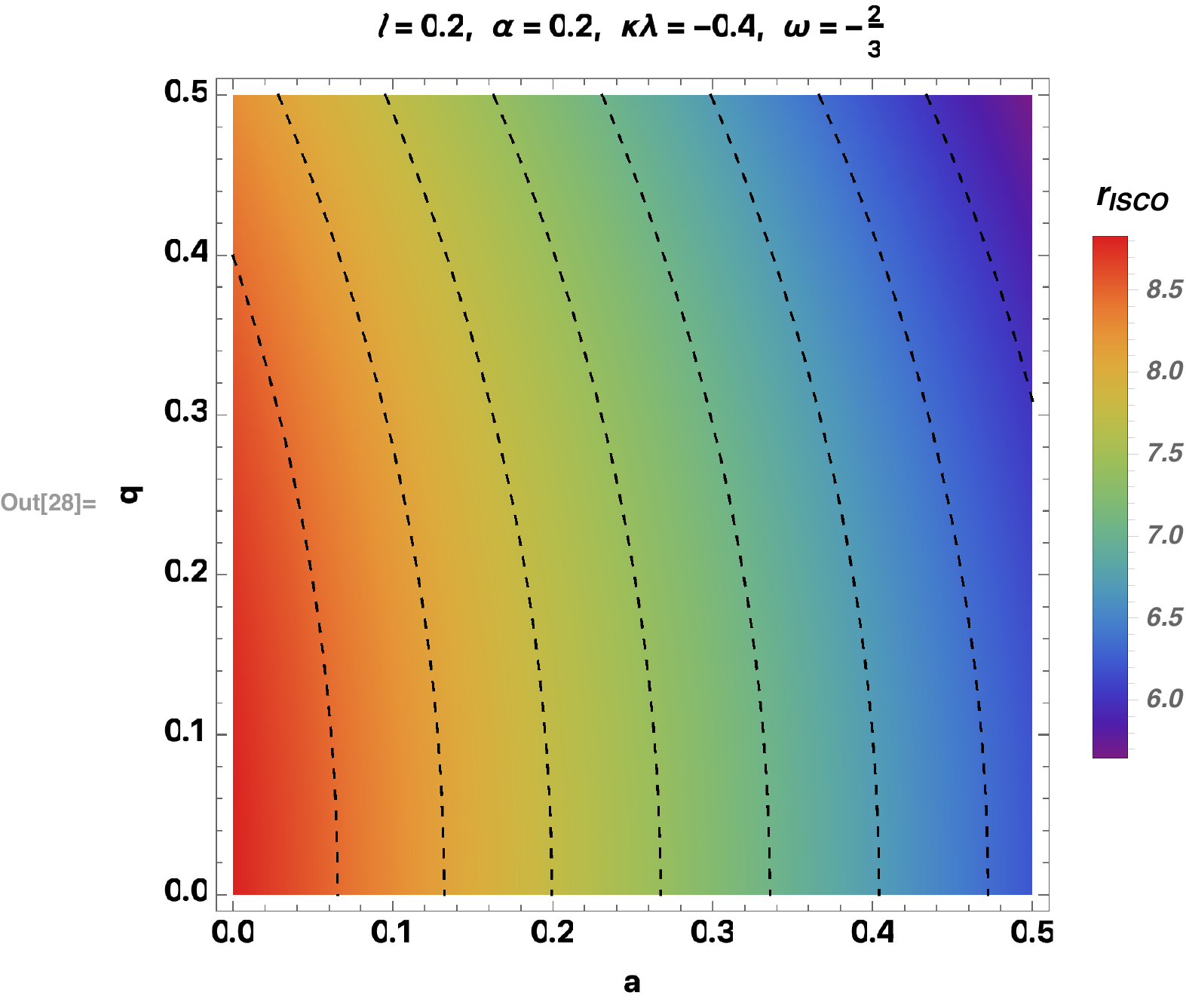}
			\includegraphics[width=0.47\linewidth]{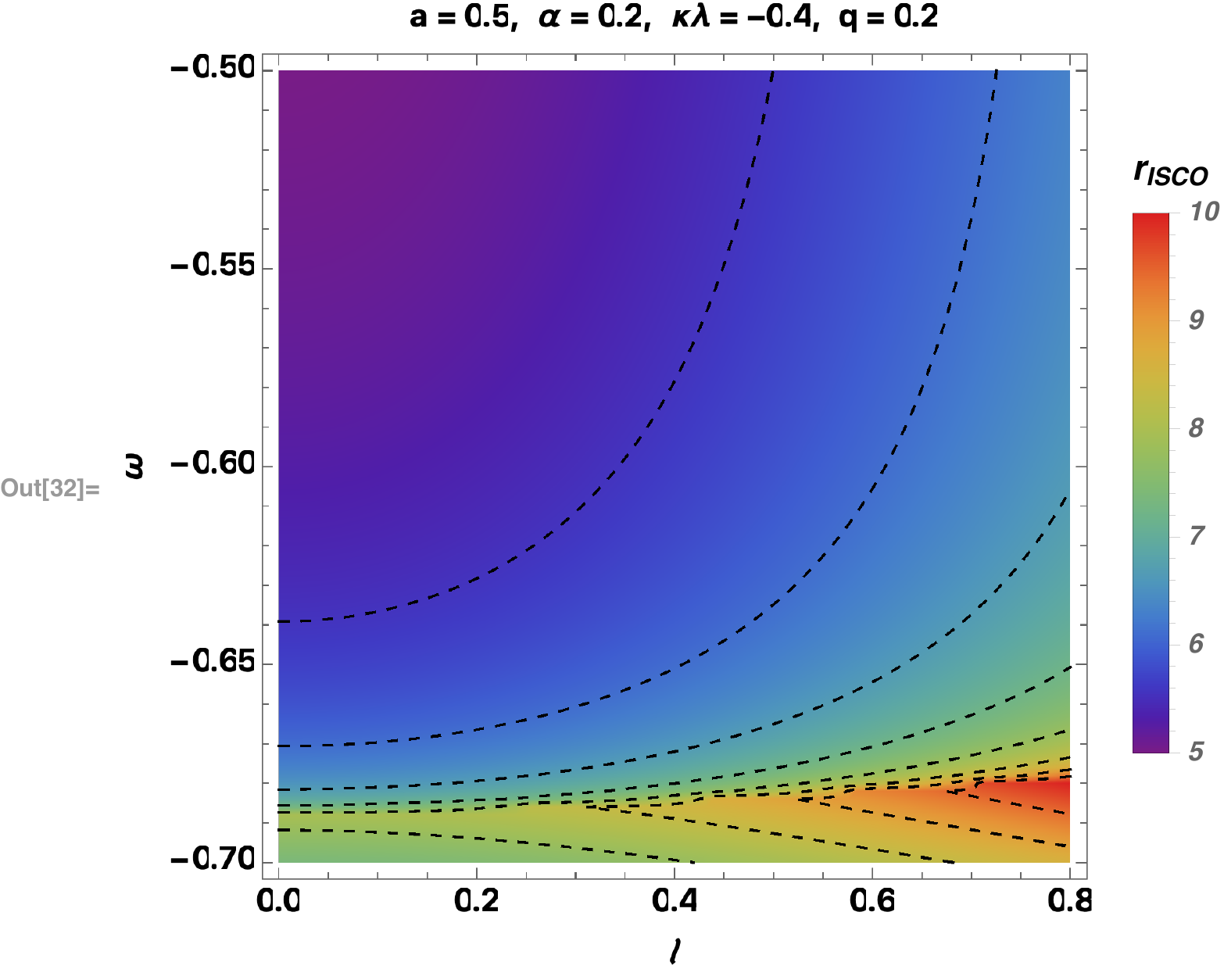}
			\includegraphics[width=0.47\linewidth]{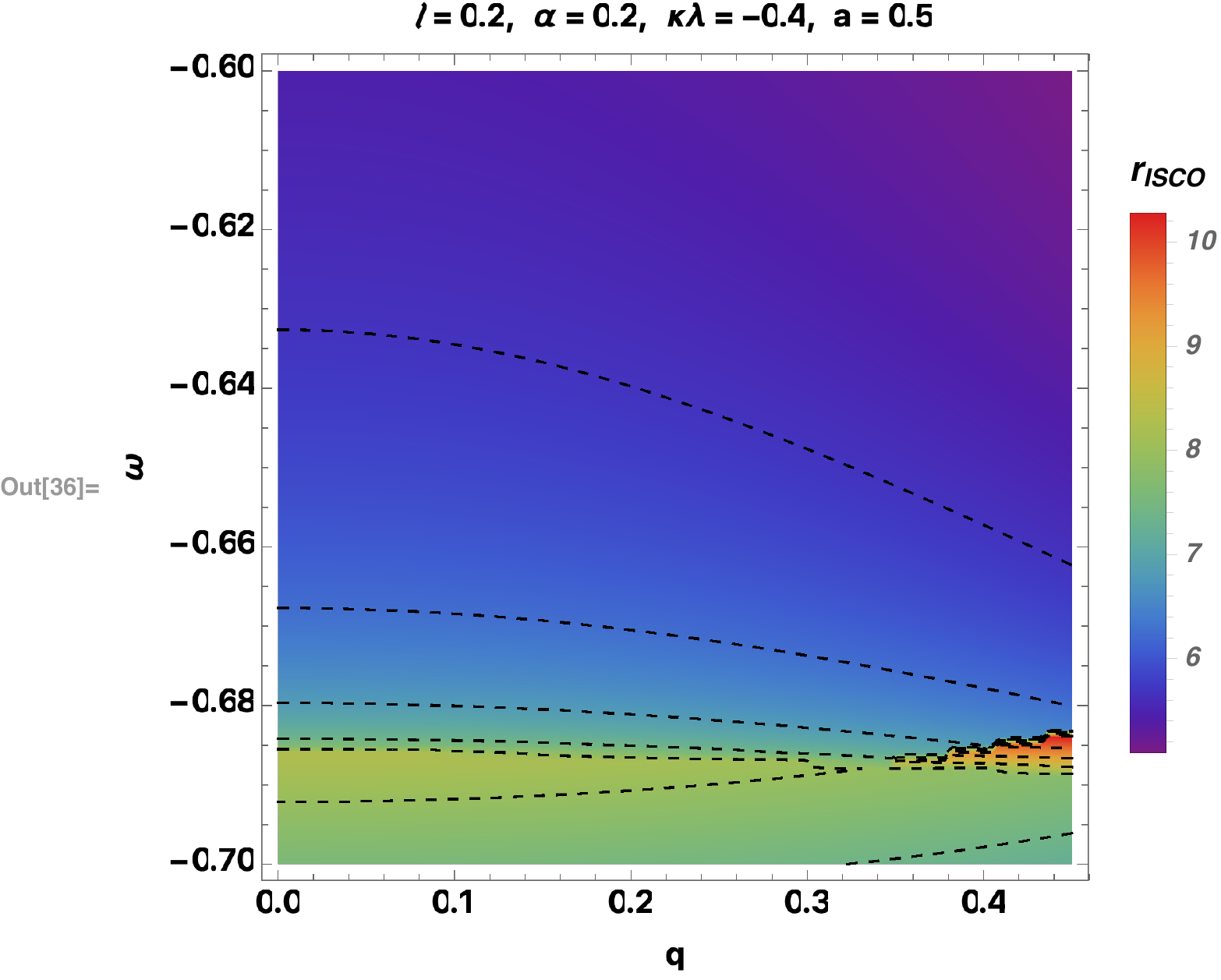}
		\end{center}
		\caption{The dependence of the ISCO radius of a test particle moving around the KNNK BH in the RG on the spacetime parameters.  \label{ISCO}}
	\end{figure*}

	With the additional condition on the effective potential that reds as 
	\begin{eqnarray}
	V_{eff}''(r)=0,\label{c3}
	\end{eqnarray}
	one can get the values of the ISCO radius of the moving test particle. One can solve the equations \eqref{c1}, \eqref{c2}, and \eqref{c3} all together numerically to get the values of the ISCO radius presented in Fig.~\ref{ISCO}. From the top-left panel of the Fig.~\ref{ISCO}, it can be illustrated for the chosen combination of the spacetime parameters that the ISCO radius becomes bigger for smaller values of the spin parameter $a$ and for bigger values of the parameter $\lambda$, reaching its maximum value around $\kappa\lambda=-0.37$. In the top-right panel of the Fig.~\ref{ISCO} it is clearly demonstrated that bigger values of the parameters $\lambda$ and $\alpha$ make the ISCO radius bigger. From the left panel of the same Fig. in the middle it is clearly demonstrated that change of the NUT parameter has very small impact on the ISCO radius of the test particle compared to the effect of the parameter $\alpha$ on it . From the right panel in the middle one can see that increase of the electromagnetic charge $q$ and the spin parameter $a$ make the ISCO radius smaller and it is also noticeable from this plot that spin parameter of the KNNK BH has stronger effect compared to the charge parameter $q$. In the bottom-left panel of the Fig.~\ref{ISCO} the change of the ISCO radius with the change of the parameters $\omega$ and $l$ is shown. From the same Fig. one can clearly see that the ISCO radius becomes bigger for the smaller values of the parameter $\omega$ and for bigger values of the NUT parameter $l$. In the bottom-right panel one can see the change of the ISCO radius with the change of the parameters $\omega$ and $q$. It is seen that the decrease of the parameter $\omega$ increases the ISCO radius and it reaches the maximum value around the point $\omega\simeq-0.69$ and $q\simeq0.5$.
	
	% \textcolor{blue}{Explanation of the figures needed to be added here}
	
	\section{QPOs of test particles near the KNNK BH in the RG}\label{sec4}
	
	 QPOs are the main observable quantities in relativistic compact objects. In order to calculate the radial and the vertical epicyclic frequencies of test particles, here we follow the standard procedure writing the  expression for the effective potential as %If one considers small perturbations of coordinates as $r=r_0+\delta r$ and $\theta=\theta_0+\delta\theta$ where $\r_0$ and $\theta_0$ correspond to the points where the effective potential takes
		\begin{eqnarray}\label{v1}
		g_{rr}\dot{r}^2+g_{\theta\theta}\dot{\theta}^2=V_{eff}.
		\end{eqnarray}
		One can separate small perturbations around circular equatorial orbits along the radial and the vertical directions in the linear weak perturbation regime.  One can assume $\dot{\theta}=0$ for the radial direction and get $\dot{r}=\dot{t}(dr/dt)$. Then the basic expression  \eqref{v1} becomes
		\begin{eqnarray}\label{v2}
		\left(\frac{dr}{dt}\right)^2=\frac{1}{g_{rr}\dot{t}^2} V_{eff}\ .
		\end{eqnarray}
		One can obtain  \eqref{v2} with respect to the coordinate $t$ in the following way
		\begin{eqnarray}
		\frac{d^2r}{dt^2}&=&\frac{1}{2} \frac{\partial}{\partial r} \left(\frac{1}{g_{rr} \dot{t}^2} V_{eff} \right)\\\nonumber
		&=& \frac{V_{eff}}{2} \frac{\partial}{\partial r} \left(\frac{1}{g_{rr} \dot{t}^2} \right)+\frac{1}{2 g_{rr} \dot{t}^2} \frac{\partial V_{eff}}{\partial r}\ .
		\end{eqnarray}
		Taking into account that $\delta r$ is a small displacement from the mean orbit as $r=r_0+\delta_r$, one can get
		\begin{eqnarray}
		\frac{d^2r}{dt^2}&=&\frac{d^2\delta_r}{dt^2}\ ,\\
		V_{eff}(r_0+\delta_r)&=&V_{eff}(r_0)+\left(\frac{\partial V_{eff}}{\partial r}\right)_{r=r_0}\delta_r\\\nonumber
		&+&O(\delta_r^2)=V_{eff}(r_0)+O(\delta_r^2)\ ,\\
		\left(\frac{\partial V_{eff}}{\partial r}\right)_{r=r_0+\delta_r}&=&\left(\frac{\partial V_{eff}}{\partial r}\right)_{r=r_0}\\\nonumber
		&+&\left(\frac{\partial^2 V_{eff}}{\partial r^2}\right)_{r=r_0} \delta_r+O(\delta_r^2)\\\nonumber
		&=&\left(\frac{\partial^2 V_{eff}}{\partial r^2}\right)_{r=r_0} \delta_r+O(\delta_r^2)\ .
		\end{eqnarray}
		In analogous way introducing a small displacement from the mean orbit $\delta_\theta$ as $\theta=\pi/2+\delta_\theta$ one can obtain expression for the vertical epicyclic frequency  with respect to the coordinate $\theta$. In the linear regime one can omit the quadratic  terms $O(\delta_r^2)$ and $O(\delta_\theta^2)$ and can get the following differential equations
		\begin{eqnarray}
		\frac{d^2\delta_r}{dt^2}+\Omega_r^2 \delta_r=0\ ,\\
		\frac{d^2\delta_\theta}{dt^2}+\Omega_r^2 \delta_\theta=0\ ,
		\end{eqnarray}
		where
		\begin{eqnarray}
		\Omega_r^2=-\frac{1}{2g_{rr}\dot{t}^2} \frac{\partial^2 V_{eff}}{\partial r^2}\ ,\\
		\Omega_\theta^2=-\frac{1}{2g_{\theta\theta}\dot{t}^2} \frac{\partial^2 V_{eff}}{\partial \theta^2}\ .
		\end{eqnarray}
	Here $\nu_r=\frac{1}{2 \pi} \frac{c^3}{G M} \Omega_r$ is the radial epicyclic frequency, $\nu_\theta=\frac{1}{2 \pi} \frac{c^3}{G M} \Omega_\theta$ is the vertical frequency, and finally $\nu_\phi=\frac{1}{2 \pi} \frac{c^3}{G M} \Omega$ is the Keplerian frequency.  %\textcolor{blue}{the text in red is taken from the Cosimo's book so, one needs to rewrite this part to avoid plagiarism}

\begin{figure*}[t!]
		\begin{center}
			\includegraphics[width=0.99\linewidth]{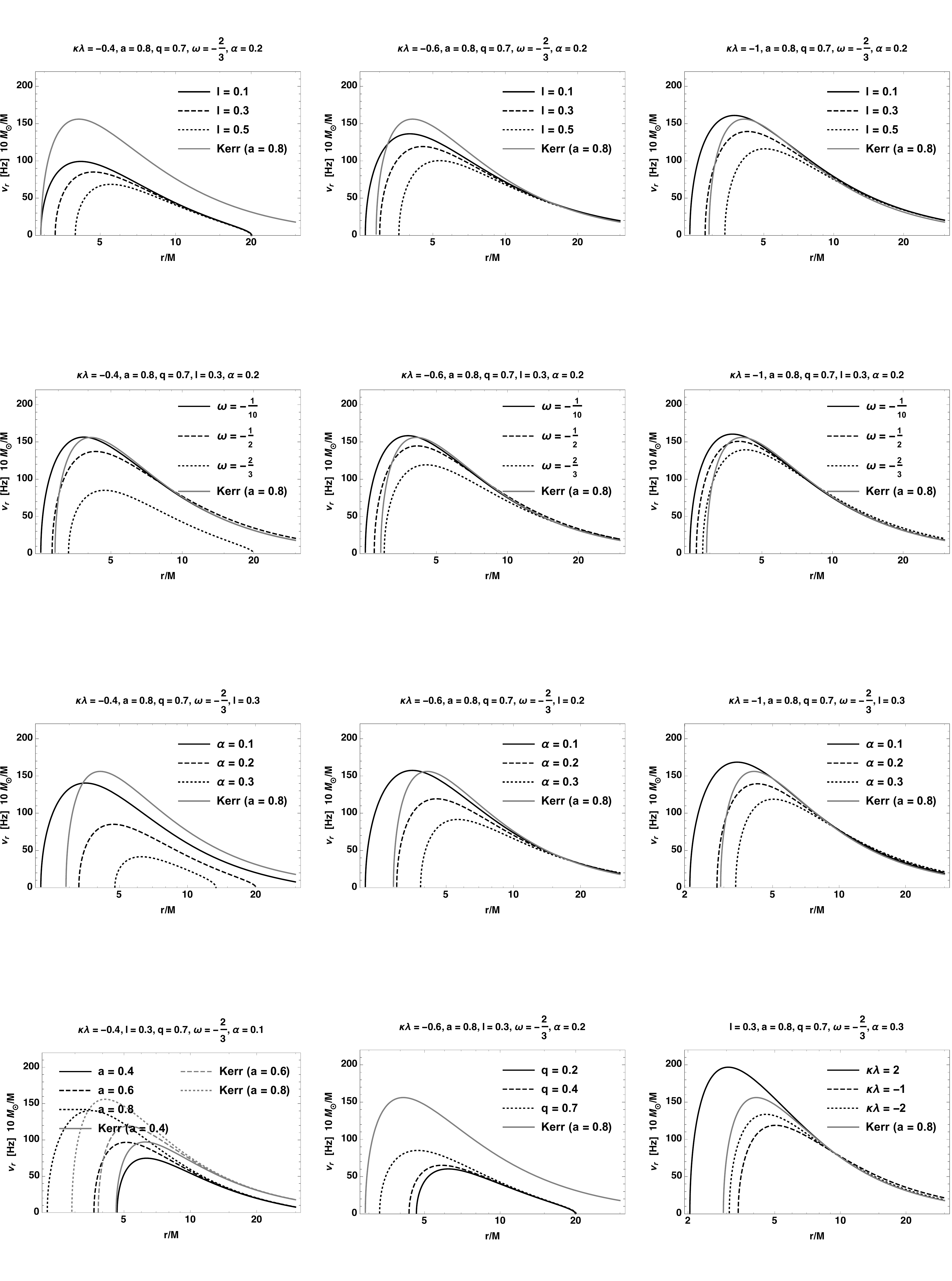}
		\end{center}
		\caption{The radial dependence of the radial epicyclic frequency, $\nu_r$ of a test particle moving around the KNNK BH in the RG. Black lines correspond to the case of KNNK BH in the RG while the gray line gives the Kerr case with the same spin parameter as the KNNK BH in the RG. \label{vr}}
	\end{figure*}

	The radial dependence of the radial epicyclic frequency is shown in Fig.~\ref{vr} in comparison with the radial frequency in the Kerr spacetime. It is clearly shown in the top panels of the Fig.~\ref{vr} that the radial epicyclic frequency becomes smaller with the increase of the NUT parameter. One can also see that decrease of the parameter $\kappa\lambda$ makes it bigger being close to the Kerr case which corresponds to the gray solid line. In the first panel it is also shown that in the case $\kappa\lambda=-0.4$ the radial epicyclic frequency becomes zero at $r\simeq20 M$. In the second row of Fig.~\ref{vr} it is shown that how the radial epicyclic frequency changes with the decrease of the parameter $\omega$ and we see that smaller frequencies correspond to smaller values of the latter. It is also interesting to see from the first panel of the second row that for the case $\omega=-\frac{2}{3}$ the radial epicyclic frequency becomes zero around the radial distance $r\simeq20 M$. Similar behaviour of the lines are shown in the first panel of the third row in the Fig.~\ref{vr}, where different values of the parameter $\alpha$ are taken. It is clearly shown that increase of the parameter $\alpha$ reduces the radial epicyclic frequency. One can also notice that the lines can double cross the abscissa when $\kappa\lambda=-0.4$. In the first panel of the last row in the Fig.~\ref{vr} the change in the radial epicyclic frequency with the change of the spin parameter $a$ is compared with radial epicyclic frequency of the Kerr BH. It is shown that for the chosen combination of the spacetime parameters the radial epicyclic frequency for the Kerr BH is always higher than that of the KNNK BH in the RG. In the second panel of the last row in the Fig.~\ref{vr} it is demonstrated that increase of the electromagnetic charge makes the radial epicyclic frequency higher. However the lines are still below the radial epicyclic frequency for the Kerr BH and become zero when $r\geq20 M$. In the last panel of the Fig.~\ref{vr} it can be seen that the change of the parameter $\lambda$ shifts the lines and it can be either higher or lower than the radial epicyclic frequency in the case of the Kerr BH.

	% \textcolor{blue}{Explanation of the figures needed to be added here}
	
\begin{figure*}[t!]
		\begin{center}
			\includegraphics[width=0.99\linewidth]{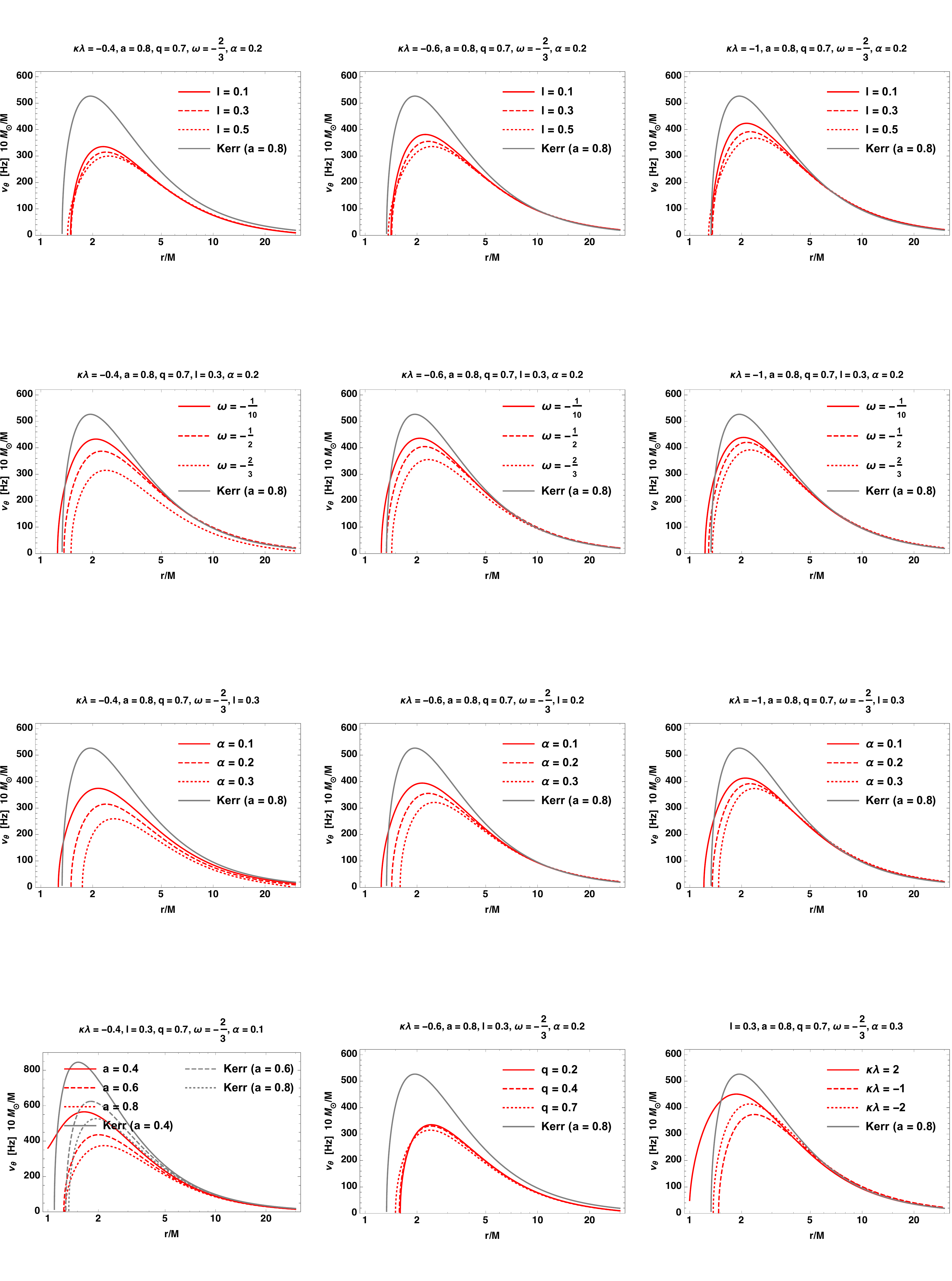}
		\end{center}
		\caption{Dependence of the vertical epicyclic frequency of the test particle, $\nu_\theta$ from the radial coordinate for the selected values of the spacetime parameters. Red lines give the values of $\nu_\theta$ for the KNNK BH in the RG case while the gray ones represent the Kerr BH case as in the previous figure for the same spin parameter of the BH. \label{vt}}
	\end{figure*}

	The radial dependence of the vertical epicyclic frequency is shown in Fig.~\ref{vt}. It is shown in the first row that change of the NUT parameter slightly changes the vertical epicyclic frequency and this change is negligible at the bigger distances. Considerable changes of the vertical epicyclic frequency are shown in the second row of Fig.~\ref{vt} for the various values of the parameter $\omega$. It is seen that reduction in this parameter reduces the vertical epicyclic frequency as well. panels in the third row of the Fig.~\ref{vt} show that the vertical epicyclic frequency becomes lower for the bigger values of the parameter $\alpha$. However, at the large distances it differs from that of the Kerr spacetime negligibly. From the first panel of the last row of the Fig.~\ref{vt} it is clearly seen that for the selected combination of the spacetime parameters the vertical epicyclic frequency of the KNNK BH in the RG is always lower than that for the Kerr BH. One can see from the second panel of the last row in the Fig.~\ref{vt} that the effect of the electromagnetic charge on the vertical epicyclic frequency is negligible. In the last panel of the same Fig. it is shown that how the increase of the parameter $\lambda$ increases the vertical epicyclic frequency. From the plots it is apparent that for the chosen combination of the spacetime parameters the vertical epicyclic frequency of the test particle moving around KNNK BH in the RG is always lower than that observed in the case of the Kerr spacetime.
	
	%\textcolor{blue}{Explanation of the figures needed to be added here}
	
\begin{figure*}[t!]
		\begin{center}
			\includegraphics[width=0.99\linewidth]{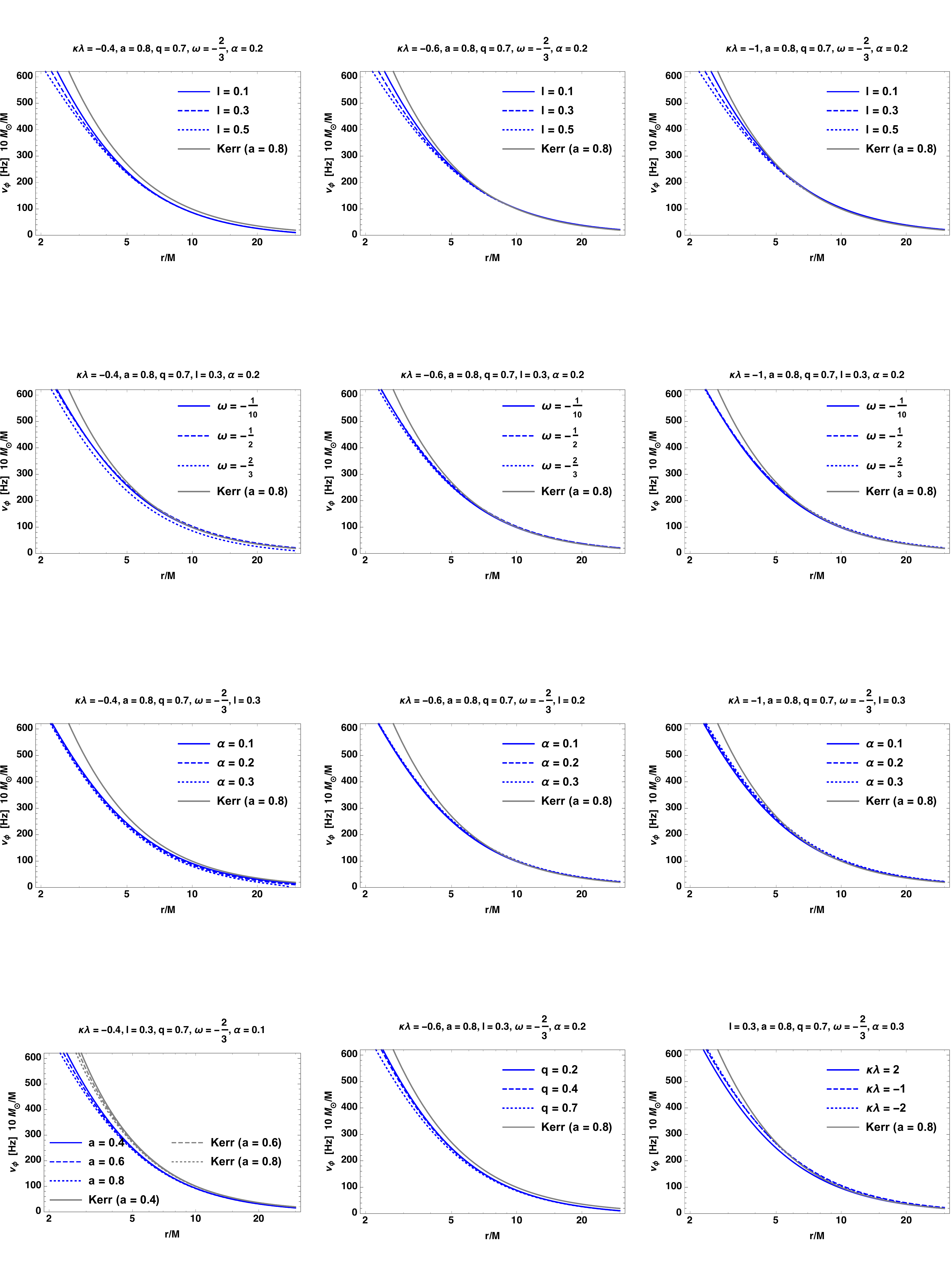}
		\end{center}
\caption{The radial dependence of the Keplerian frequency of a test particle, $\nu_\phi$ for the different values of the spacetime parameters. Blue lines are for the KNNK BH in the RG case and the gray ones for the Kerr BH with the same spin parameter. \label{vp}}
	\end{figure*}

	The radial dependence of the Keplerian frequency is shown in Fig.~\ref{vp}. One can clearly see from the plots presented in this Fig. that the behavior of the lines for the Keplerian frequency is almost identical with the pure Kerr BH spacetime and only small departures from the latter can be detected in the far regions. One can conclude from the results of the Fig.~\ref{vp} that influence of the spacetime parameters of the KNNK BH in the RG on the Keplerian frequency of a test particle is very small and almost the same as in the Kerr spacetime at the large distances and only in the vicinity of the KNNK BH in RG, the differences can become considerable.
	
	%\textcolor{blue}{Explanation of the figures needed to be added here}
	
	\section{Conclusion}\label{sec5}
In this work test particle motion around KNNK BH in the RG has been investigated together with the fundamental frequencies of test particles to probe the spacetime around such a BH. The effective potential of the test particle moving in the equatorial plane of the KNNK BH in the RG has been evaluated. It has been shown that increase of the parameters $l$ and $\alpha$ increases the effective potential, while the increase of the parameters $\omega$, $a$, and $q$ reduces it. ISCO radius of test particles has also been calculated and it has been demonstrated for the selected combinations of the spacetime parameters that increase of the parameters $l$, $\alpha$ and $\lambda$ increase the ISCO radius while increase in the rest of the parameters of the KNNK spacetime in RG, decrease it. However, there were several exceptions in the regions of the KNNK spacetime in the RG, presented in the figures where ISCO radius takes its maximum value and then starts going down with the change of some parameters as discussed in the main text above.\\ 

Further we have investigated the fundamental frequencies of test particle near the circular orbits in the vicinity of the KNNK BH in the RG. It has been shown that change of the parameter $\lambda$ can make the radial epicyclic frequency to become zero at some larger distances from the central source, for the chosen combinations of the spacetime parameters. Investigation of the vertical epicyclic frequency has shown that for the same values of the spin of the KNNK BH in the RG and the Kerr BH, the latter always has higher vertical epicyclic frequency. Investigation of the Keplerian frequency in the case of KNNK BH in the RG, however, has shown that there are small deviations from the Kerr case in the far regions from the central BH and only in the regions near to the BH one can observe considerable differences of such a frequency.

It should be noted that the explicit expressions for the energy, angular momentum, effective potential, and fundamental frequencies calculated for KNNK BH in the RG are very complicated and too lengthy to fit in the paper and we have presented them graphically.	
	
	\section*{Acknowledgments}
	B.N. acknowledges support from the China Scholarship Council (CSC), grant No. 2018DFH009013. A.A. and B.A. acknowledge the support of Uzbekistan Ministry for Innovative Development Grants and  the Abdus Salam International Centre for Theoretical Physics under the Grant No.  OEA-NT-01. 

	\section*{Data Availability Statement}
	Data sharing not applicable to this article as no datasets were generated or analyzed during the current study.

	\bibliographystyle{apsrev4-1}
\bibliography{gravreferences}

\end{document}